\documentclass[fleqn]{article}
\usepackage{fleqn}
\usepackage{graphics}
\usepackage{epsfig}
\usepackage{amsmath}
\usepackage{amssymb}
\usepackage{calc}
\newcommand{\beq}{\begin{equation}}
\newcommand{\eeq}{\end{equation}}
\newcommand{\bea}{\begin{eqnarray}}
\newcommand{\eea}{\end{eqnarray}}
\begin{document}
\Large
\begin{center}
{\bf  A fermionic code related to the exceptional group $E_8$}
\end{center}
\large
\vspace*{-.1cm}
\begin{center}
P\'eter L\'evay$^{1,2}$ 
and Fr\'ed\'eric Holweck$^{1}$
 \end{center}
 \vspace*{-.4cm} \normalsize
 \begin{center}

$^{1}$Laboratoire Interdisciplinaire Carnot de Bourgogne, ICB/UTBM, UMR 6303 CNRS,
Universit\'e Bourgogne Franche-Comt\'e, 90010 Belfort Cedex, France

$^{2}$Department of Theoretical Physics, Institute of Physics, Budapest University of\\
 Technology and Economics and MTA-BME Condensed Matter Research Group, H-1521 Budapest, Hungary

\vspace*{.0cm}
\vspace*{.2cm} (19 January 2018)
\end{center}
\vspace*{-.3cm} \noindent \hrulefill

\vspace*{.1cm} \noindent {\bf Abstract:}
In this paper we study the Hamming-like fermionic code encoding three-qubits into sixteen Majorana modes recently introduced by Hastings.
We show that although this fermionic code cannot be obtained from a single qubit stabilizer code via the usual procedure however, it can be obtained from two, qubit stabilizer ones via a glueing procedure combining both single and double occupancy embeddings of qubits into fermionic Fock space.
This technique identifies the code subspace as a Cartan subspace of the largest exceptional group $E_8$. 
In arriving at these results we develop a general formalism for implementing this glueing procedure via the use of intertwiners between different subsectors of the Fock space realizing embedded qubits.

\vspace*{.3cm} 
\\ \hspace*{1.95cm} 
 \vspace*{-.2cm} \noindent \hrulefill

\section{Introduction}

It is well-known that from qubit stabilizer codes\cite{Gott,Rains,Nielsen} one can construct Majorana fermion codes\cite{Terhal}.
Namely: given an $n$ qubit stabilizer code encoding $k$ logical qubits one can construct a Majorana fermion code\cite{Kitaev} with $4n$ Majorana modes encoding $k$ qubits. It is also known that if the distance of the qubit stabilizer code is $d$ then the distance for the corresponding fermionic one is $2d$.
An elementary example of a distance $2$ qubit stabilizer code is featuring $4$ physical qubits and $2$ logical ones. Let us call this code the "four qubit code".
It is easy to see
that there is no distance $2$ qubit stabilizer code on $4$ physical qubits and $3$ logical ones\cite{Hastings}. Hence according to the fermionic embedding procedure no code with $16$ Majorana modes, distance $4$, and $3$ logical qubits exists.
However, in a recent paper Hastings\cite{Hastings} managed to construct a class of Majorana fermionic codes with $2^l$ modes 
$l\geq 3$, which \emph{cannot be obtained from a qubit stabilizer code}. 
The $l=4$ case of $16$ modes will be of special importance for us. This code is precisely a Majorana fermionic one with $16$ Majorana modes distance $4$ and $3$ logical qubits i.e. a $[16,3,4]$ code.

In this paper we show that although this special code cannot be obtained from a \emph{single} qubit stabilizer one\cite{Hastings} however, it can be obtained via combining \emph{two} (qubit stabilizer) codes in a special manner. 
As a first step one can start with \emph{two copies} of the four qubit code.
One of the copies is based on an embedded four qubit system in the single and the other on an embedded four qubit one in the double occupancy representation\cite{LH}.
Then the two copies are glued together to form Hasting's code via applying the intertwiner between the single and double occupancy representations. This intertwiner acts as the bit flip operation acting on an extra qubit, i.e. the third one, making it possible to combine the two copies of (logical) two qubit systems to a (logical) three qubit one.
As an extra bonus this construction identifies this code as a Cartan subspace of the largest exceptional group $E_8$.

The aim of this paper is two-fold. Apart from being a case study aiming at the group theoretical clarification of the meaning of a special code we would like to set the stage for further understanding of fermionic codes within the formalism as developed in Ref.\cite{LH}.
Our approach of concentrating on the code subspaces rather than on their stabilizers is somewhat contrary to the spirit of the stabilizer formalism. Although lacking the main virtue and elegance of the stabilizer formalism our observations hopefully provide some hints for constructing physically interesting Majorana fermion codes which cannot be obtained from qubit stabilizer ones.

The organization of this paper is as follows.
In Section 2. we summarize the basics needed for understanding fermionic codes. Section 3. is devoted to a recapitulation of ideas presented in Ref.\cite{LH} on embedding qubits into fermionic Fock space.
Here the notions of single and double occupancy representations are introduced.
In Section 4. the simple example of the four qubit code is considered from the dual perspective of regarding this code as a $[16,2,4]$ fermionic one both in the single and the double occupancy representations.
In Section 5. it is shown that the $[16,3,4]$ code introduced by Hastings can be regarded as a one glued together from two copies of the $[16,2,4]$ four qubit code. 
Finally we link our construction to the Cartan decomposition $\mathfrak{e}_8=\mathfrak{so}(16)\oplus \mathfrak{m}$ of the Lie algebra of the largest exceptional group $E_8$ where $\mathfrak{m}$ is the even chirality sector of the fermionic Fock space with $8$ modes. Section 6. is left for the conclusions.

\section{Fermionic codes based on an even number of modes}

As a starting point we summarize our setting up for fermionic systems and their codes.
Let $V$ be an $N=2n$ dimensional complex vector space and $V^{\ast}$ its dual.
We regard $V={\mathbb {C}}^N$ with $\{e_{I}\}, I=1,2,\dots N$ the canonical basis and $\{e^{I}\}$ the dual basis. Elements of $V$ will be called {\it single particle states}. To the  number $N=2n$ we will refer to as the number of {\it modes}\footnote{The number of \emph{modes} ($2n$) is to be contrasted with the number of \emph{Majorana modes} ($4n$) to be introduced soon.}.
Equipped with a Hermitian inner product $\langle\cdot\vert\cdot\rangle$ our vector space $V$ can be regarded as the Hilbert space of single particle states.
We also introduce the $2N$ dimensional vector space
\beq
\mathcal {V}\equiv V\oplus V^{\ast}
\label{nu}.
\eeq
\noindent
An element of $\mathcal{V}$ is of the form $x=v+\alpha$ where $v$ is a vector
and $\alpha$ is a linear form.
According to the method of second quantization to any element $x\in\mathcal{V}$ one can associate a linear operator $\mathcal{O}_x$ acting on the fermionic Fock space $\mathcal{F}_{N}$ which is
represented as a direct sum of $m$-particle subspaces $\mathcal{F}^{(m)}, m=0,1,\dots N$.

Explicitely, to the basis vectors of $\mathcal{V}$ one associates fermionic creation and annihilation operators
\beq
\mathcal{O}_{{e}^{I}}\equiv{p}_I,\qquad \mathcal{O}_{{e}_{J}}\equiv{n}_{J},\qquad I,J=1,\dots N
\label{defioper}
\eeq
\noindent
satisfying the usual
fermionic anticommutation relations \beq
\{{p}_{I},{n}_{J}\}=\delta_{IJ},\qquad
\{{p}_{I},{p}_{J}\}=\{{n}_{I},{n}_{J}\}=0.
\label{anticommute} \eeq \noindent 
Since a Hermitian inner product on $V$ is at our disposal one can regard the ${p}_{I}$ as the Hermitian conjugate of the ${n}_{I}$.
In this way one can revert to the usual notation familiar from the literature: $f_I^{\dagger}\equiv p_I$ and $f_I\equiv n_I$.
Then the direct sum structure of $\mathcal{F}_N$  is made explicit as follows.
Define the vacuum state  $\vert\rm{vac}\rangle\in\mathcal{F}_N$ by the property
\beq
{n}_{I}\vert{\rm vac}\rangle =0.
\label{vacuum}
\eeq
\noindent
Then $\mathcal{F}_{N}$ of dimension $2^N$ is spanned by the basis vectors
\beq
\vert {\rm vac}\rangle,\qquad
p_{I_1}{p}_{I_2}\cdots{p}_{I_m}\vert{\rm vac} \rangle,\qquad 1\leq I_1<I_2<\cdots<I_m\leq N, \qquad m=1,\dots N.
\label{fockbasis}
\eeq
Hence
\beq
\mathcal{F}_{N}=\bigoplus_{m=0}^N\mathcal{F}^{(m)}.
\label{bigo}
\eeq
\noindent

There is an alternative way of looking at the structure of $\mathcal{F}_{N}$.
In this picture one describes the $2^N$ basis vectors of $\mathcal{F}_{N}$ as
\beq
\vert \kappa_1\kappa_2\dots\kappa_N\rangle \equiv p_1^{\kappa_1}p_2^{\kappa_2}\cdots p_N^{\kappa_N}\vert{\rm vac}\rangle,\qquad (\kappa_1,\kappa_2,\dots,\kappa_N)\in {{\mathbb Z}_2}^N
\label{betolt}
\eeq
\noindent
hence $\vert 00\dots 0\rangle\equiv\vert{\rm vac}\rangle$, $\vert 10\dots 0\rangle\equiv p_1\vert{\rm vac}\rangle$ etc.
Let us now define $2N$ Majorana fermion operators as follows
\beq
{c}_{2I-1}={p}_I+{n}_I,\qquad {c}_{2I}=i({p}_I-{n}_I).
\label{maj1}
\eeq
\noindent
These operators are satisfying 
\beq
\{c_{\mu},c_{\nu}\}=2\delta_{\mu\nu},\qquad c_{\mu}^{\dagger}=c_{\mu},\qquad \mu=1,2,\dots 2N.
\label{maj2}
\eeq
\noindent
Representing the $c_{\mu}$ via the $N$-fold tensor product\footnote{Only in the following formula for the product the use of the $\otimes$ symbol is implicit.} of Pauli matrices (with identity matrices in the remaining slots implicit) as
\beq
c_{2I-1}=\left(\prod_{J=1}^{I-1}\sigma_z^{(J)}\right)\sigma_x^{(I)}\equiv 
\prod_{J=1}^{I-1}Z_JX_I
,\qquad
c_{2I}=\left(\prod_{J=1}^{I-1}\sigma_z^{(J)}\right)\sigma_y^{(I)}\equiv \prod_{J=1}^{I-1}Z_JY_I
\label{JW}
\eeq
\noindent
amounts to the well-known Jordan-Wigner representation. Hence in this alternative picture $\mathcal{F}_{N}$ is represented as an $N$-fold tensor product of two dimensional complex vector spaces.

Define the chirality operator as
\beq
\Gamma=\prod_{I=1}^N(1-2p_In_I)=(-i)^N\prod_{I=1}^{N}c_{2I-1}c_{2I}
\label{chirop}
\eeq
\noindent
For $k$ even (odd) the states of Eq.(\ref{fockbasis}) are eigenvectors of $\Gamma$ with eigenvalues $+1 (-1)$.
The corresponding eigensubspaces of $\Gamma$ will be denoted by ${\mathcal F}_N^{\pm}$.
We will refer to their elements as spinors of positive or negative chirality.

Now we summarize the basics of fermionic codes\cite{Terhal}.
The formalism follows the same pattern as the one familiar for stabilizer codes\cite{Nielsen}.
The single-mode operators $c_1,\dots, c _{2N}$,
together with the phase factor $i$, generate a group of Majorana operators $\it{Maj}(2N)$.
The number $2N=4n$ will be called the number of \emph{Majorana modes}.
An element of the group $\it{Maj}(2N)$ can be represented as 
\beq
\omega c_{\mathcal{A}},\qquad 
c_{\mathcal{A}}=\prod_{\mu\in\mathcal{A}}c_\mu,\qquad
\omega\in\{\pm 1,\pm i\}
\label{omega}
\eeq
\noindent
where $\mathcal{A}\subset \{1,\dots 2N\}$ is
some subset of modes.

We use a standard ordering for the product of single-mode operators $c_{\mu}$, meaning
that the indices $\mu$ increase from the left to the right. The weight of a Majorana
operator is the number of modes in its support, $\vert c_{\mathcal A}\vert = \vert\mathcal{A}\vert$. A Majorana operator
is called even (odd) iff its weight is even (odd).
We have
\beq 
c_{\mathcal{A}}c_{\mathcal{B}}=(-1)^{\langle \mathcal{A},\mathcal{B}\rangle}c_{\mathcal{B}} c_{\mathcal{A}}
,\qquad
\langle \mathcal{A},\mathcal{B}\rangle\equiv \vert\mathcal{A}\vert\cdot\vert\mathcal{B}\vert+\vert\mathcal{A}\cap\mathcal{B}\vert\quad {\rm mod}2.
\label{szimpi}
\eeq
\noindent
Regarding the power set of $\{1,\dots 2N\}$ with $2^{2N}$ elements as the vector space $\mathbb{Z}_2^{2N}$ over $\mathbb{Z}_2$ with addition defined by the symmetric difference of two subsets, $\langle\cdot\vert\cdot\rangle$ defines a symplectic form on $\mathbb{Z}_2^{2N}$. For the basic properties of these structures we orient the reader to our Appendix.

A {Majorana fermion code is a linear subspace of $\mathcal{F}_N$ which is left invariant by a \emph{stabilizer group} $S\subset \it{Maj}(2N)$
satisfying the conditions:
1. $S$ is an Abelian group not containing $-1$,
2. all elements of  $S$ have even weight.

The set of Majorana operators $C(S)$ that commute with all elements
of $S$ is called the centralizer of
$S$. Logical operators
of a Majorana fermion code associated to $S$ are elements of $C(S)$ which are not in $S$. If
$S$ is generated by $N-k$ independent generators, then $C(S)$ is generated by $N+k$
independent ones. One can choose a set of $2k$ logical Pauli operators $\{\overline{X}_1,\dots ,\overline{X}_k,
\overline{Z}_1,\dots ,\overline{Z}_k\}\in C(S)-S$
 satisfying the usual Pauli commutation relations.
The distance $d$ of a Majorana fermion code is the minimum weight of logical operators.
A code with distance $d$ is able to detect any error affecting
less than $d$ Majorana modes, and able to correct any error acting on less than $d/2$ Majorana modes. A fermionic code with $2N=4n$ Majorana modes, $k$ logical bits and distance $d$ will be called an $[4n,k,d]$
fermionic code. Clearly codes with $d=2$, though legitime ones, are not particularly useful since they cannot correct the detected error on a single Majorana mode. However, they are important as building blocks for nontrivial fermionic codes. Examples of codes of that type are the ones related to embedding qubits in Fock space. We will discuss these codes in the next section.

\section{Embedding qubits}

In this section based on the formalism of \cite{LH} we summarize results on embedding qubits into fermionic Fock space.
Being of basic importance for our concerns, the single and double occupancy representations will be carefully discussed.

According to (\ref{bigo}) an arbitrary state $\vert\psi\rangle\in\mathcal{F}_N$ can be written in the
form \beq 
\vert\psi\rangle=\sum_{m=0}^N\sum_{I_1I_2\cdots
I_m=1}^N\frac{1}{m!}\psi^{(m)}_{I_1I_2\dots
I_k}{p}_{I_1}{p}_{I_2}\cdots {p}_{I_m}\vert{\rm vac}\rangle. \label{steexp}
\eeq \noindent Here the $m$th order totally antisymmetric tensors
$\psi^{(m)}_{I_1I_2\cdots I_m}$ encapsulate the complex amplitudes
of the $m$-"particle" subspace.
Let us define
\beq
{s}\equiv\frac{1}{2}{A_{IJ}}[{p}_I,{n}_J]+\frac{1}{2}B_{IJ}{p}_I{p}_J+\frac{1}{2}C_{IJ}{n}_I{n}_J, 
\label{sexplicit} \eeq \noindent
then $\Lambda\equiv e^{{s}}\in{\rm
Spin}(2N,\mathbb{C})$. 
Here $A,B,C$ are $N\times N$ complex matrices with $B$ and $C$ skew-symmetric and for the repeated indices summation is understood.
Transformations of the form \beq
\vert\psi\rangle\mapsto \lambda
\Lambda\vert\psi\rangle,\qquad
(\lambda,\Lambda)\in\mathbb{C}^{\times}\times {\rm
Spin}(2N,\mathbb{C}),\qquad
\vert\psi\rangle\in\mathcal{F}_N \label{genslocc} \eeq
\noindent are called generalized SLOCC transformations \cite{LH,SL}.
In the following we are interested in the 
$\rm{Spin}(2N,\mathbb{C})$ subgroup represented by elements of the form $(1,\Lambda)$.

The particle number conserving subgroup of the
generalized SLOCC group is obtained by setting $B=C=0$ in
Eq.(\ref{sexplicit}).
Now we have \beq
\Lambda=e^{-{\rm
Tr}A/2}e^{A_{IJ}{p}_I{n}_J}. \label{valodislocc} \eeq
\noindent
The action of $\Lambda$ on a state of the
$m$-particle subspace $\mathcal{F}^{(m)}$ of $\mathcal{F}_N$ is
\beq \Lambda\vert{\psi}^{(m)}\rangle=\frac{1}{m!}\psi^{\prime
(m)}_{I_1\dots I_m} {p}_{I_1}\cdots {p}_{I_m}\vert
\rm{\rm vac}\rangle,\quad \psi^{\prime (m)}_{I_1\dots I_m}
=({\rm{Det}\mathcal{A}})^{-1/2}{\mathcal{A}_{J_1}}^{I_1}\cdots
{\mathcal{A}_{J_k}}^{I_k} \psi^{(k)}_{I_1\dots I_k}
\label{fertraf} \eeq \noindent 
where
\beq
{\mathcal{A}}=e^{{A}}\in GL(N,\mathbb{C}).
\eeq
\noindent

Recall that for the cases considered here the number of modes is even, $N=2n$.
In this case we introduce
a new ("odd-even") labelling for the canonical basis vectors of $V$ as \beq
\{e_1, e_{\overline{1}},e_2,e_{\overline{2}},\dots ,e_n,e_{\overline{n}}\}\equiv\{e_1,e_2,e_3,e_4,\dots,e_{2n-1},e_{2n}\}. \label{ujracim} \eeq \noindent 
Hence we have two sets of basis vectors $\{e_j\}$ and $\{e_{\overline{j}}\}$ where  $j=1,\dots n$.
For the states of the $n$-fermion subspace $\mathcal{F}^{(n)}$ of $\mathcal{F}_{2n}$ we introduce the special notation \beq \vert
Z\rangle=\frac{1}{n!}Z_{I_1I_2\dots
I_n}{p}_{I_1}{p}_{I_2}\cdots {p}_{I_n}\vert{\rm vac} \rangle.
\label{Z} \eeq \noindent Under the subgroup of
Eq.(\ref{valodislocc})
the amplitudes of this state transform
as\beq Z_{J_1\dots J_n}\mapsto
({\rm{Det}\mathcal{A}})^{-1/2}{\mathcal{A}_{J_1}}^{I_1}\cdots
{\mathcal{A}_{J_n }}^{I_n} {Z}_{I_1\dots I_n}\equiv
{S_{J_1}}^{I_1}\cdots {S_{J_n }}^{I_n} {Z}_{I_1\dots I_n},
\label{fertrafn} \eeq \noindent where \beq {S_J}^I\equiv ({\rm
Det}\mathcal{A})^{-\frac{1}{2n}}{\mathcal{A}_J}^I\in
SL(2n,\mathbb{C}). \label{sl2n} \eeq \noindent Hence in this
special case the  (\ref{valodislocc}) subgroup of transformations
coming from the group ${\rm Spin}(4n,\mathbb{C})$ with
$B_{IJ}=C_{IJ}=0$ will produce an $SL(2n,\mathbb{C})$ subgroup.

From the set of $2n\choose n$ basis vectors of  $\mathcal{F}^{(n)}$ we choose a special subset containing merely $2^n$ elements as follows
\beq
{p}_1{p}_2\cdots{p}_n\vert {\rm vac}\rangle,\quad
{p}_1{p}_2\cdots{p}_{\overline{n}}\vert{\rm vac}\rangle,\quad\dots,\quad
{p}_{\overline{1}}{p}_{\overline{2}}\cdots{p}_n\vert {\rm vac}\rangle,\quad {p}_{\overline{1}}{p}_{\overline{2}}\cdots {p}_{\overline{n}}\vert{\rm vac} \rangle,
\label{nqubsub}
\eeq
\noindent
or in the notation of Eqs.(\ref{betolt}) and (\ref{ujracim}), 
\beq
\vert 10 10 \dots 10\rangle,\qquad \vert 1010\dots 01\rangle,\quad\dots,\quad\vert 0101\dots 10\rangle,\qquad
\vert 0101\dots 01\rangle
\eeq
\noindent
These basis vectors are spanning the linear subspace ${\mathcal K}_s$ of an embedded $n$-qubit system. 
We will refer to this embedding of $n$-qubits as the \emph{ single occupancy} representation\cite{LH,Cheng}.

We give the explicit form of this embedding as follows.
For the $n$-qubit state $\vert\psi\rangle\in \mathbb{C}^{2^n}$ with amplitudes $\psi_{00\dots 0}$,
$\psi_{00\dots 1}\dots$, $\psi_{11\dots 1}$.
we associate an element $\vert Z_{\psi}\rangle\in {\mathcal{K}}_s\subset\mathcal{F}^{(n)}$ via the mapping
\beq
\vert\psi\rangle\mapsto \vert Z_{\psi}\rangle=
\left(\psi_{11\dots 1}{p}_1{p}_2\cdots{p}_n+
\psi_{11\dots 0}
{p}_1{p}_2\cdots{p}_{\overline{n}}+\cdots +
\psi_{00\dots 0}
{p}_{\overline{1}}{p}_{\overline{2}}\cdots{p}_{\overline{n}}
\right)\vert {\rm vac}\rangle.
\label{pricminmap}
\eeq
\noindent

Now the operators\footnote{The overline on $\sigma_s$ $s=x,y,z$ is \emph{not} referring to complex conjugation. It is merely indicating that the Pauli matrices obtained in this way are different from the ones showing up in Eq.(\ref{JW}).}
\beq
{\overline{\sigma}_x}^{(j)}=p_jn_{\overline{j}}+p_{\overline{j}}n_j,\quad
{\overline{\sigma}_y}^{(j)}=i(p_jn_{\overline{j}}-p_{\overline{j}}n_j),\quad
{\overline{\sigma}_z}^{(j)}=p_{\overline{j}}n_{\overline{j}}-p_{j}n_j
\label{Kitaev}
\eeq
taken together with the triple $\{i{\overline{\sigma}_x}^{(j)},i{\overline{\sigma}_y}^{(j)},i{\overline{\sigma}_z}^{(j)}\}$ form the $2\times 2$ representation for the generators of $n$ copies of the group $SL(2,\mathbb{C})$.
Clearly on the basis states (the computational basis for the j-th qubit)  
\beq
\vert {\overline 0}\rangle_j\equiv \dots p_{\overline{j}}\dots\vert{\rm vac}\rangle= \vert \dots 01\dots\rangle,\qquad
\vert {\overline 1}\rangle_j\equiv \dots p_j\dots\vert{\rm vac}\rangle= \vert \dots 10\dots\rangle
\label{singleembed}
\eeq
\noindent 
these operators give rise to the usual Pauli spin matrices.
Then the $n$ copies of $SL(2,\mathbb{C})$s give rise to an  $SL(2,\mathbb{C})^{\times n}$ action on $\mathcal{K}_s$.

One can relate this action to a special subset of the fermionic $SL(2n,\mathbb{C})$ transformations on $\mathcal{F}^{(n)}$ 
of the (\ref{fertrafn}) form
as follows. These are
transformations, characterized by a $2n\times 2n$ matrix
$S$ of the (\ref{sl2n}) form, that leave the subspace $\mathcal{K}_s$ invariant.
Looking at Eq.(\ref{fertrafn}) it is easy to see that such
transformations can be organized to a matrix of the form \beq
{S}=\begin{pmatrix}a&b\\c&d\end{pmatrix}\in SL(2n,\mathbb{C})
\label{nq1} \eeq \noindent where $a=diag(a_1,\dots a_n)$,
$b=diag(b_1,\dots b_n)$, $c=diag(c_1,\dots c_n)$,
$diag(d_1,\dots d_n)$, i.e. the $n\times n$ blocks of $S$ are
{\it diagonal matrices}. One can also place these complex numbers
into an $n$ element set of $2\times 2$ matrices \beq
S^{(j)}\equiv\begin{pmatrix}a_j&b_j\\c_j&d_j\end{pmatrix}\in
SL(2,\mathbb{C}),\quad j=1,2,\dots n. \eeq \noindent 
Now the transformation $\vert
Z_{\psi}\rangle\mapsto \Lambda\vert Z_{\psi}\rangle$ gives
rise to the one \beq \psi_{\alpha_1\dots\alpha_n}\mapsto
{{{S}^{(1)}}_{\alpha_1}}^{\beta_1} \cdots
{{{S}^{(n)}}_{\alpha_n}}^{\beta_n}\psi_{\beta_1\dots\beta_n},\qquad
{S}^{(j)}\in SL(2,\mathbb{C}),\quad j=1,2\dots n \eeq
\noindent which is just the $SL(2,\mathbb{C})^{\times n}$ action for $n$-qubits.

The single occupancy representation can be elegantly described in terms of majorana operators $c_{\mu}, \mu=1,2,\dots 4n$ introduced in Eqs. (\ref{maj1})-(\ref{maj2}).
On $\mathcal{F}_{2n}$ we have
\beq
\overline{\sigma}_x^{(j)}=\frac{i}{2}(c_{4j-3}c_{4j}-c_{4j-2}c_{4j-1}),
\label{Kitaev2a}
\eeq
\noindent
\beq
\overline{\sigma}_y^{(j)}=\frac{i}{2}(c_{4j-2}c_{4j}-c_{4j-1}c_{4j-3}),
\label{Kitaev2b}
\eeq
\noindent
\beq
\overline{\sigma}_z^{(j)}=\frac{i}{2}(c_{4j-1}c_{4j}-c_{4j-3}c_{4j-2}).
\label{Kitaev2c}
\eeq
\noindent
However, on $\mathcal{K}_s$ we have $c_{4j-3}c_{4j}=-c_{4j-2}c_{4j-1}$ etc. hence
on $\mathcal{K}_s$ one can choose
\beq
\overline{\sigma}_x^{(j)}=ic_{4j-3}c_{4j},\quad
\overline{\sigma}_y^{(j)}=ic_{4j-2}c_{4j},\quad
\overline{\sigma}_z^{(j)}=ic_{4j-1}c_{4j},\qquad j=1,2,\dots n.
\label{Kitaev3}
\eeq
\noindent
This result connects our single occupancy embedding of qubits into fermions to the one of Kitaev\cite{Kitaev}.

Let us now introduce the operators
\beq
g_j\equiv c_{4j-3}c_{4j-2}c_{4j-1}c_{4j},\qquad j=1,2,\dots n
\label{geoperatorok}
\eeq
\noindent
satisfying
\beq
[g_j,g_k]=0,\qquad j,k=1,2,\dots n.
\label{komm}
\eeq
\noindent
These operators stabilize $\mathcal{K}_s$.
In the language of fermionic stabilizer codes one can regard $\mathcal{K}_s$ as a fermionic code, with stabilizer
group defined by the generators $g_j$.
Indeed, the dimension of $\mathcal{F}_{2n}$ is $2^{2n}=2^{N_{\rm maj}/2}$ where $N_{\rm maj}=4n$ is the number of Majorana modes, on the other hand since the number of generators is $n$, we have for the dimension\cite{Hastings} of $\mathcal{F}^{(n)}$
$2^{2n}/2^n=2^n$ the dimension of the $n$-qubit Hilbert space.
In this language the single occupancy subspace is a $[4n,n,2]$ fermionic code.

Apart from the single occupancy embedding for $n$-qubits there exist the double and mixed occupancy embeddings\cite{LH}. The \emph{double occupancy} subspace ${\mathcal{K}}_d$ is spanned by the basis states 
\beq
\vert{\rm vac}\rangle,\quad p_jp_{\overline{j}}\vert{\rm vac}\rangle,\quad
p_{j}p_{\overline{j}}p_{k}p_{\overline{k}}\vert{\rm vac}\rangle \quad j<k,\quad\dots,\quad
p_{1}p_{\overline{1}}\cdots p_{n}p_{\overline{n}}\vert{\rm vac}\rangle.
\eeq
\noindent
For our $n$-qubit state $\vert\psi\rangle$  now we associate a new state $\vert W_{\psi}\rangle$, called a one in the double occupancy representation, by the rule 
\beq
\vert W_{\psi}\rangle=\psi_{00\dots 0}p_{1}p_{\overline{1}}
p_{2}p_{\overline{2}}
\cdots p_{n}p_{\overline{n}}\vert{\rm vac}\rangle
+\cdots +\psi_{01\dots 1}p_1p_{\overline{1}}\vert{\rm vac}\rangle +\psi_{11\dots 1}\vert{\rm vac}\rangle.
\label{doccup}
\eeq
\noindent
Now the operators
\beq
\tilde{\sigma}_x^{(j)}=n_{\overline{j}}n_j-p_{\overline{j}}p_j,\quad
\tilde{\sigma}_y^{(j)}=i(n_{\overline{j}}n_j+p_{\overline{j}}p_j),\quad
\tilde{\sigma}_z^{(j)}=p_{\overline{j}}n_{\overline{j}}-n_{j}p_j
\label{Kitaev4}
\eeq
taken together with the triple $\{i\tilde{\sigma}_x^{(j)},i\tilde{\sigma}_y^{(j)},i\tilde{\sigma}_z^{(j)}\}$ form the $2\times 2$ representation for the generators\cite{LH} of $n$ copies of the group $SL(2,\mathbb{C})$.
Clearly on the basis states (the computational basis for the j-th qubit)
\beq
\vert \tilde{0}\rangle_j\equiv \dots p_jp_{\overline{j}}\dots\vert{\rm vac}\rangle=\vert \dots 11\dots\rangle,\qquad
\vert \tilde{1}\rangle_j\equiv\dots {\mathbf 1}\dots\vert\rm{vac}\rangle=\vert\dots 00\dots\rangle
\label{doubleembed}
\eeq
\noindent
these operators again give rise to the usual Pauli spin matrices.

The double occupancy form of the operators (\ref{Kitaev3}) is
\beq
{\tilde{\sigma}_x}^{(j)}=-\frac{i}{2}(c_{4j-3}c_{4j}+c_{4j-2}c_{4j-1}),
\label{Kitaev5a}
\eeq
\noindent
\beq
{\tilde{\sigma}_y}^{(j)}=\frac{i}{2}(c_{4j-2}c_{4j}+c_{4j-1}c_{4j-3}),
\label{Kitaev5b}
\eeq
\noindent
\beq
{\tilde{\sigma}_z}^{(j)}=\frac{i}{2}(c_{4j-1}c_{4j}+c_{4j-3}c_{4j-2}).
\label{Kitaev5c}
\eeq
\noindent
It is easy to relate the single and double occupancy subspaces\cite{LH}.
Define the intertwiner
\beq
\Omega\equiv c_1c_5\cdots c_{4n-3}
\label{Omega}
\eeq
\noindent
then we have
\beq
{\mathcal{K}}_d
=
\Omega{\mathcal{K}}_s
\label{intertwine}
\eeq
\noindent
and 
\beq
{\tilde{\sigma}_r}^{(j)}=c_{4j-3}\overline{\sigma}_s^{(j)}c_{4j-3},\qquad r=x,y,z,\qquad j=1,2,\dots ,n.
\label{relate}
\eeq
\noindent

From Eqs.(\ref{Kitaev2a})-(\ref{Kitaev2c}) and Eqs.(\ref{Kitaev5a})-(\ref{Kitaev5c}) 
and the (\ref{singleembed}), (\ref{doubleembed}) definitions it is clear
that the operators
\beq
\hat{\sigma}_x^{(j)}\equiv -ic_{4j-2}c_{4j-1},\qquad
\hat{\sigma}_y^{(j)}\equiv ic_{4j-2}c_{4j},\qquad
\hat{\sigma}_z^{(j)}\equiv i c_{4j-1}c_{4j}
\label{same}
\eeq
\noindent
form a representations of the Pauli operators on \emph{both the single and double occupancy} subspaces.

\section{Embeddings of the four qubit code}

The four qubit code is a $[4,2,2]$ qubit stabilizer code with with $n=4$, $k=2$, and $d=2$ i.e. it is a distance\footnote{
Recall\cite{Terhal} that the weight of a Pauli operator is the number of qubits on which the operator acts nontrivially.
The distance of the code is the minimal weight of its logical operators.}
two code which is encoding two logical qubits into four ones. Using again the shorthand notation $(\sigma_x,\sigma_y,\sigma_z)=(X,Y,Z)$ the stabilizer group of this code is $S=\langle XXXX,ZZZZ\rangle$.
The centralizer of the stabilizer is
\beq
C(S)=\langle XIIX,IXIX,ZIIZ,IZIZ,S,iI\rangle.
\label{stab4qcode}
\eeq
\noindent
The two logical qubits are encoded into the subspace $V_S$ spanned by the basis vectors
\beq
\overline{\vert 00\rangle}=\vert 0000\rangle +\vert 1111\rangle,
\quad
\overline{\vert 01\rangle}=\vert 0101\rangle +\vert 1010\rangle,
\label{egy}
\eeq
\beq
\overline{\vert 10\rangle}=\vert 1001\rangle +\vert 0110\rangle,
\quad
\overline{\vert 11\rangle}=\vert 0011\rangle +\vert 1100\rangle.
\label{ketto}
\eeq
We remark that states of $V_S$ 
of the form
\beq
\vert G_{abcd}\rangle=a\overline{\vert 00\rangle}+b\overline{\vert 01\rangle}+c\overline{\vert 10\rangle}+d\overline{\vert 11\rangle},\quad a,b,c,d,\in{\mathbb C}\label{Gabcd}
\eeq are comprising the semisimple SLOCC orbits of four-qubit entangled states\cite{Verstraete,Djokovic}. 

The $15$ nontrivial logical operations of the two-qubit Pauli group acting on the encoded qubits can be constructed from the basis
 \beq
\{\overline{IX},\overline{XI}, \overline{IZ},\overline{ZI}\}\leftrightarrow \{XIXI,XIIX,ZIIZ,ZIZI\}.
\label{logicalops}
\eeq
Note that on $V_S$, $IIXX\simeq XXII$ etc. since $IIXX=XXII\cdot XXXX$ where $XXXX\in S$.
Indeed, due to $\overline{IZ}\cdot\overline{XI}=\overline{XZ}$ and $ZIIZ\cdot XIIX=-YIIY$ one should have $-\overline{XZ}\leftrightarrow YIIY$ etc.
Hence for example the following famous arrangement of nine two-qubit Pauli operators (Mermin square\cite{Mermin}) can be represented by four-qubit ones acting on $V_S$ as follows
\beq
\begin{pmatrix}\overline{XX}&\overline{IX}&\overline{XI}\\\overline{YY}&-\overline{ZX}&-\overline{XZ}\\
\overline{ZZ}&\overline{ZI}&\overline{IZ}\end{pmatrix}\leftrightarrow
\begin{pmatrix}{XXII}&{XIXI}&{XIIX}\\{YYII}&{YIYI}&{YIIY}\\
{ZZII}&{ZIZI}&{ZIIZ}\end{pmatrix}
\label{merminke}
\eeq
Here the operators on each row and column are mutually commuting however, the product of the rows is the identity with a \emph{plus} and the product of the columns is the identity with a \emph{minus} sign.
($XXII\cdot XIXI\cdot XIIX=XXXX\simeq IIII$ on $V_S$.)
Notice also that all the logical operators showing up in Eq.(\ref{merminke}) are of weight two in accordance with the fact that our code has distance two.

Untill this point the two qubits were embedded into the four-qubit space.
But using the fermionic formalism makes it also possible to embed our two-qubits into a bigger, majorana fermionic system with sixteen modes.
Indeed, this embedding is effected by regarding the four-qubit Hilbert space as a fermionic system with eight modes
in the single occupancy representation.
In this case we have $n=4$ and $N=8$, the generalized SLOCC
group is $\mathbb{C}^{\times}\times{\rm Spin}(16,\mathbb{C})$, and the single occupancy representation is embedded inside the space ${\mathcal F}_8^+$, i.e. the one of
Weyl spinors of positive chirality. Such positive chirality spinors are of the form \beq
\vert\Psi_+\rangle=\left(\eta+\frac{1}{2!}{\rm X}_{IJ}{p}_{IJ}+\frac{1}{4!}{\rm Z}_{IJKL}{p}_{IJKL}+
\frac{1}{6!}\varepsilon_{IJKLMNRS}{\rm Y}_{IJ}{p}_{KLMNRS}+
\xi{p}_{12345678} \right)\vert{\rm vac} \rangle \label{alap}\eeq
\noindent 
where we have used the notation $p_{I_1}p_{I_2}\cdots p_{I_k}\equiv p_{I_1I_2\dots I_k}$. According to Eq.(\ref{pricminmap}) in the single occupancy representation a four qubit state $\vert\psi\rangle$
is embedded into $\mathcal{F}^{(4)}\subset\mathcal{F}^+_8\subset\mathcal{F}_8$ as follows \beq
\vert {\rm Z}_{\psi}\rangle=\left({\rm Z}_{1234}{p}_{1234}+{\rm Z}_{123\overline{4}}
{p}_{123\overline{4}}+\cdots+{\rm Z}_{\overline{123}4}p_{\overline{123}4}+{\rm Z}_{\overline{1234}}
{p}_{\overline{1234}}\right)\vert{\rm vac}
\rangle\in\mathcal{K}_s,\label{4qssss} \eeq \noindent
\beq
{\rm Z}_{1234}=\psi_{1111},\quad {\rm Z}_{123\overline{4}}=\psi_{1110},\quad\dots\quad,{\rm Z}_{\overline{123}4}=\psi_{0001},\quad 
{\rm Z}_{\overline{1234}}=\psi_{0000}
\eeq
\noindent
i.e. from the $128$ amplitudes of the spinor $\Psi_+$ only $16$ amplitudes, taken from the ${8\choose 4}=70$ ones of ${\rm Z}_{IJKL}$, are kept. 

The single occupancy subspace $\mathcal{K}_s\subset\mathcal{F}^{(4)}$ as a Majorana fermionic code can be uniquely characterized by the stabilizer group $S_4\subset {\rm Maj}(16)$ given as $S_4=\langle g_1,g_2,g_3,g_4\rangle$ where the generators are the $n=4$ ones given by Eqs.(\ref{geoperatorok})-(\ref{komm}). 
The $[4,2,2]$ four qubit code as a stabilizer code reinterpreted as a $[16,2,4]$ Majorana fermionic one is obtained by adjoining two extra generators $g_5$ and $g_6$ explicitely given by
\beq
g_5=c_1c_4c_5c_8c_9c_{12}c_{13}c_{16},\qquad
g_6=c_3c_4c_7c_8c_{11}c_{12}c_{15}c_{16}.
\label{x4z4}
\eeq
\noindent
Indeed by virtue of Eq.(\ref{Kitaev3}) $g_5$ and $g_6$ correspond to $\overline{\sigma}_x^{(1)}\overline{\sigma}_x^{(2)}\overline{\sigma}_x^{(3)}\overline{\sigma}_x^{(4)}$ and $\overline{\sigma}_z^{(1)}\overline{\sigma}_z^{(2)}\overline{\sigma}_z^{(3)}\overline{\sigma}_z^{(4)}$ respectively, corresponding to the two generators of $S$ introduced in the beginning of this section.
Let us denote our new stabilizer by $S_2\subset {\rm Maj}(16)$ where $S_2=\langle g_1,g_2,g_3,g_4,g_5,g_6\rangle$.
Hence in the notation of Section 2. this code is an $[16,2,4]$ fermionic one\cite{Terhal}. 
Notice that $S_2$ automatically contains the chirality operator $\Gamma=g_1g_2g_3g_4=\prod_{I=1}^8c_{2I-1}c_{2I}$. This corresponds to the fact that the code space of Eq.(\ref{Gabcd}) is embedded into ${\mathcal F}_8^+$.

Let us denote by $\overline{V}_{S_2}$ the code space stabilized by $S_2$.
Then an arbitrary element of the code space can be written in the following form
\beq
\vert\overline{G}\rangle=\left[a(p_{1234}+p_{\overline{1234}})+b(p_{1\overline{2}3\overline{4}}+p_{\overline{1}2\overline{3}4})
+c(p_{1\overline{23}4}+p_{\overline{1}23\overline{4}})+d(p_{12\overline{34}}+p_{\overline{12}34})
\right]\vert{\rm vac}\rangle
\label{fermi2qcode}
\eeq
\noindent
which is the embedded version of the state of Eq.(\ref{Gabcd}) into fermionic Fock space in the single occupancy representation.

For later use it is worth recapitulating the meaning of the basis vectors of $\overline{V}_{S_2}$.
Define
\beq
\vert \overline{E}_0\rangle =(p_{1234}+p_{\overline{1234}})\vert\rm{vac}\rangle =\vert 01010101\rangle +\vert 10101010\rangle=\vert\overline{0000}\rangle +\vert\overline{1111}\rangle =\vert\overline{\overline{00}}\rangle
\label{1aa}
\eeq
\noindent
\beq
\vert \overline{E}_1\rangle =(p_{1\overline{2}3\overline{4}}+p_{\overline{1}2\overline{3}4})\vert\rm{vac}\rangle =\vert 01100110\rangle +\vert 10011001\rangle =\vert\overline{0101}\rangle +\vert\overline{1010}\rangle = \vert\overline{\overline{01}}\rangle
\eeq
\noindent
\beq
\vert \overline{E}_2\rangle =(p_{1\overline{23}4}+p_{\overline{1}23\overline{4}})\vert\rm{vac}\rangle =\vert 01101001\rangle +\vert 10010110\rangle =\vert\overline{0110}\rangle +\vert\overline{1001}\rangle =\vert\overline{\overline{10}}\rangle
\eeq
\noindent
\beq
\vert \overline{E}_3\rangle =(p_{12\overline{34}}+p_{\overline{12}34})\vert\rm{vac}\rangle =\vert 01011010\rangle +\vert 10100101\rangle
=\vert\overline{0011}\rangle +\vert\overline{1100}\rangle =\vert\overline{\overline{11}}\rangle
\label{1dd}
\eeq
\noindent
then we can write
\beq
\vert\overline{G}\rangle =\overline{y}_{\alpha}\vert \overline{E}_{\alpha}\rangle,\qquad (\overline{y}_0,\overline{y}_1,\overline{y}_2,\overline{y}_3)=(a,b,c,d),\qquad \alpha=0,1,2,3. 
\eeq
\noindent
The meaning of the alternative forms for $\vert E_{\alpha}\rangle$ is as follows. The appearance of the vectors showing up after the first two equality signs conforms with the definitions displayed by Eqs.(\ref{betolt}) and (\ref{ujracim}).
The form displayed after the third equality sign conforms with our (\ref{singleembed}) realization of the single occupancy representation.
The overline in this context refers to the fact that this representation corresponds to a fermionic code with the overlined bits being its logical bits. The last form of the basis vectors indicates that these vectors span the code subspace of the four qubit code living inside the single occupancy subspace. The double overline refers to the fact that it can be regarded as a "code inside a code". Of course apart form the new notation dictated by the fermionic context the corresponding vectors are just the ones of Eqs.(\ref{egy})-(\ref{ketto}).  

The four qubit code can easily be realized inside the double occupancy subspace as well.
In order to do this one simply has to use the intertwiner of Eq.(\ref{intertwine}) namely
\beq
\Omega \equiv c_1c_5c_9c_{13}
\label{Omega4}
\eeq
\noindent
to create the corresponding basis vectors spanning the relevant code subspace $\tilde{V}_{S_2}$
\beq
\vert\tilde{E}_{\alpha}\rangle\equiv \Omega\vert\overline{E}_{\alpha}\rangle.
\label{atmegy}
\eeq
\noindent
The result is the state
\beq
\vert\tilde{G}\rangle =\tilde{y}_{\alpha}\vert \tilde{E}_{\alpha}\rangle,\qquad (\tilde{y}_0,\tilde{y}_1,\tilde{y}_2,\tilde{y}_3)=(a,b,c,d),\qquad \alpha=0,1,2,3.
\eeq
\noindent
with
\beq
\vert \tilde{E}_0\rangle =({\mathbf 1}+p_{1234\overline{1234}})\vert\rm{vac}\rangle =\vert 00000000\rangle +\vert 11111111\rangle=\vert{\tilde{1}\tilde{1}\tilde{1}\tilde{1}}\rangle +\vert{\tilde{0}\tilde{0}\tilde{0}\tilde{0}}\rangle =\vert{\tilde{\tilde{0}}\tilde{\tilde{0}}}\rangle
\eeq
\noindent
\beq
\vert \tilde{E}_1\rangle =(p_{1\overline{1}3\overline{3}}+p_{2\overline{2}4\overline{4}})\vert\rm{vac}\rangle =\vert 11001100\rangle +\vert 00110011\rangle =\vert{\tilde{0}\tilde{1}\tilde{0}\tilde{1}}\rangle +\vert{\tilde{1}\tilde{0}\tilde{1}\tilde{0}}\rangle = \vert{\tilde{\tilde{0}}\tilde{\tilde{1}}}\rangle
\eeq
\noindent
\beq
\vert \tilde{E}_2\rangle =(p_{1\overline{1}4\overline{4}}+p_{2\overline{2}3\overline{3}})\vert\rm{vac}\rangle =\vert 11000011\rangle +\vert 00111100\rangle =\vert{\tilde{0}\tilde{1}\tilde{1}\tilde{0}}\rangle +\vert{\tilde{1}\tilde{0}\tilde{0}\tilde{1}}\rangle =\vert{\tilde{\tilde{1}}\tilde{\tilde{0}}}\rangle
\eeq
\noindent
\beq
\vert \tilde{E}_3\rangle =(p_{1\overline{1}2\overline{2}}+p_{3\overline{3}4\overline{4}})\vert\rm{vac}\rangle =\vert 11110000\rangle +\vert 00001111\rangle
=\vert{\tilde{0}\tilde{0}\tilde{1}\tilde{1}}\rangle +\vert{\tilde{1}\tilde{1}\tilde{0}\tilde{0}}\rangle =\vert{\tilde{\tilde{1}}\tilde{\tilde{1}}}\rangle
\eeq
\noindent
Now the form displayed after the third equality sign conforms with our (\ref{doubleembed}) realization of the double occupancy representation.

Notice that according to Eqs.(\ref{same}), and (\ref{logicalops}) the two qubit logical operators of the four qubit code acting \emph{ on both the single and double} occupancy subspaces are of the form
\beq
\{\overline{\overline{IX}},\overline{\overline{XI}},\overline{\overline{IZ}},\overline{\overline{ZI}}\}=
\{-c_2c_3c_{10}c_{11},-c_2c_3c_{14}c_{15},-c_3c_4c_{15}c_{16},-c_3c_4c_{11}c_{12}\}
\label{ezitt1}
\eeq
\noindent
\beq
\{\tilde{\tilde{I}}\tilde{\tilde{X}},\tilde{\tilde{X}}\tilde{\tilde{I}},\tilde{\tilde{I}}\tilde{\tilde{Z}},\tilde{\tilde{Z}}\tilde{\tilde{I}}\}=
\{-c_2c_3c_{10}c_{11},-c_2c_3c_{14}c_{15},-c_3c_4c_{15}c_{16},-c_3c_4c_{11}c_{12}\}.
\label{ezitt2}
\eeq
\noindent

\section{The fermionic code of Hastings}

The fermionic code of Hastings is a $[16,3,4]$ fermionic code with the following stabilizer group $S_3\subset {\rm Maj}(16)$.
\beq
S_3=\langle G_1, G_2, G_3, G_4, \Gamma\rangle,\qquad G_j=\prod_{\mu=1}^{16}(c_{\mu})^{v_j^{\mu}},\qquad v_j=v_j^{\mu}e_{\mu}\in {\mathbb Z}_2^{16},
\label{Hastings}
\eeq
\noindent
where the components of the four vectors $v_j$ with $j=1,2,3,4$ are described by the rows of the following $4\times 16$ matrix

\setcounter{MaxMatrixCols}{16}
\beq
\begin{bmatrix}0 & 1 & 0 & 1 & 0 & 1 & 0 & 1 & 0 & 1 & 0 & 1 & 0 & 1 & 0 & 1 \\
               0 & 0 & 1 & 1 & 0 & 0 & 1 & 1 & 0 & 0 & 1 & 1 & 0 & 0 & 1 & 1 \\
	       0&0&0&0&1&1&1&1&0&0&0&0&1&1&1&1\\
               0&0&0&0&0&0&0&0&1&1&1&1&1&1&1&1
	       \end{bmatrix}.
\label{komp}
\eeq
\noindent
Alternatively (see Appendix\footnote{
In Eq.(\ref{formula}) the rows of (\ref{komp}) should be regarded as \emph{columns}.
}) one can describe these generators as the $8$-fold Kronecker products of Pauli matrices
\beq
G_1=XYXYXYXY,\quad G_2=IZIZIZIZ,\quad G_3= IIZZIIZZ,\quad G_4=IIIIZZZZ. 
\eeq
\noindent
Notice also that 
$G_5\equiv\Gamma=ZZZZZZZZ$.

From Eq.(\ref{komp}) one can see that the $16$ columns of the matrix give the binary representation of integers from $0$ to $15$. Hence any error effected by the action of a \emph{single} Majorana operator $c_{\mu}$ gives a unique error syndrome.
Indeed, the pattern of violated stabilizers is given by the binary representation of $\mu-1$. Clearly the construction of this code is closely related to a similar one based on the classical Hamming code\cite{Hastings}.

Now we show that although this code cannot be obtained from a single qubit stabilizer code\cite{Hastings} via the usual procedure\cite{Terhal} however, one can obtain it from \emph{two} such stabilizer ones via a simple glueing procedure based on a use of the single and double occupancy representations. 
Take two different copies of the code subspace of the $[4,2,2]$ four qubit code.
Inside the Fock space these qubit stabilizer codes give rise to two code subspaces of $[16,2,4]$ fermionic codes.
One ($\overline{V}_{S_2}$) with amplitudes $(\Psi_0,\Psi_1,\Psi_2,\Psi_3)$ belonging to the single and the other ($\tilde{V}_{S_2}$) with amplitudes $(\Psi_4,\Psi_5,\Psi_6,\Psi_7)$ to the double occupancy representation.
Glueing them together with the intertwiner $\Omega$ of Eq.(\ref{Omega4}) we obtain the state
\beq
\vert\Psi\rangle= \Psi_{\alpha}\vert \overline{E}_{\alpha}\rangle +\Psi_{\alpha +4}\Omega\vert \overline{E}_{\alpha}\rangle\qquad \alpha=0,1,2,3
\label{psi3}
\eeq
\noindent
where for the definitions see Eqs.(\ref{1aa})-(\ref{1dd}) and (\ref{atmegy}).
We claim that the basis vectors showing up in this state are spanning $\overline{\overline{V}}_{S_3}$, i.e. the code space of the $[16,3,4]$ fermionic code stabilized by $S_3$.

In order to show this one just has to observe that the generators of $S_3$ stabilize all of the basis vectors. 
In order to find a convenient set of logical operators we convert the decimal labels of the amplitudes of $\vert\Psi\rangle$ to a three-qubit binary one by using the following assignment for the basis vectors
\beq
\vert\overline{E}_0\rangle =\vert\overline{\overline{000}}\rangle,\qquad
\vert\overline{E}_1\rangle =\vert\overline{\overline{001}}\rangle,\qquad
\vert\overline{E}_2\rangle =\vert\overline{\overline{010}}\rangle,\qquad
\vert\overline{E}_3\rangle =\vert\overline{\overline{011}}\rangle
\eeq
\noindent
\beq
\Omega\vert\overline{E}_0\rangle =\vert\overline{\overline{100}}\rangle,\qquad
\Omega\vert\overline{E}_1\rangle =\vert\overline{\overline{101}}\rangle,\qquad
\Omega\vert\overline{E}_2\rangle =\vert\overline{\overline{110}}\rangle,\qquad
\Omega\vert\overline{E}_3\rangle =\vert\overline{\overline{111}}\rangle.
\eeq
\noindent
Labelling the three logical qubits from the left to the right, flipping the first qubit amounts to going from the single occupancy representation to the double and vice versa. Hence the intertwiner $\Omega=c_1c_5c_9c_{13}=YZXIYZXI$ serves as the logical Pauli operator $\overline{\overline{XII}}$.
Notice also that any of the operators $g_j$ with $j=1,2,3,4$ of Eq.(\ref{geoperatorok}) 
stabilize the basis vectors of the single occupancy 
representation and taking to the negative of the basis vectors of the double occupancy one. 
Hence any one of them can serve as the logical Pauli operator $\overline{\overline{ZII}}$. 
As a representative of this operator let us choose $g_1=c_1c_2c_3c_4=-ZZIIIIII$.
The remaining set of logical operators acting on the second and third logical qubit are immediately identified by looking at 
Eqs.(\ref{ezitt1})-(\ref{ezitt2}). Indeed, since the very same operators are acting on the single and double occupancy spaces one has the assignment
\beq
\{\overline{\overline{IIX}},\overline{\overline{IXI}},\overline{\overline{IIZ}},\overline{\overline{IZI}}\}=
\{-c_2c_3c_{10}c_{11},-c_2c_3c_{14}c_{15},-c_3c_4c_{15}c_{16},-c_3c_4c_{11}c_{12}\}.
\label{ezitt3}
\eeq
\noindent
Hence the centralizer of the stabilizer $S_3$ is
\beq
C(S)=\langle -c_2c_3c_{10}c_{11},-c_2c_3c_{14}c_{15}, c_1c_5c_9c_{13},-c_3c_4c_{15}c_{16},-c_3c_4c_{11}c_{12},
c_1c_2c_3c_4, i{\mathbf 1},S\rangle.
\eeq
\noindent

In order to elucidate the group theoretical meaning of the code of Hastings we proceed as follows.
First of all observe that the Lie algebra of the $Spin(16,\mathbb{C})$ subgroup of the (\ref{genslocc}) generalized SLOCC group is $\mathfrak{spin}(16)\simeq\mathfrak{so}(16)$. This Lie algebra has ${16\choose 2}=120$ generators.
The $Spin(16,\mathbb{C})$ subgroup acts on the $2^7=128$ dimensional space of positive chirality spinors $\mathcal{F}_8^+$ in a manner as described by Eq.(\ref{genslocc}). Let us denote this $128$ dimensional vector space by the symbol $\mathfrak{m}$.
Taken the direct sum of this vector space with the $120$ dimensional vector space coming from the Lie algebra of $\mathfrak{so}(16)$ gives rise to a $248$ dimensional one of the form $\mathfrak{so}(16)\oplus\mathfrak{m}$.
One can give a Lie-algebra structure to this vector space as follows.
The $\mathfrak{so}(16)$ part is already a Lie algebra, we regard this as a Lie subalgebra. The infinitesimal version of the generalized SLOCC action defines the commutator of the $\mathfrak{so}(16)$ part with the $\mathfrak{m}$ one. 
The only set of commutators yet to be defined are the $[\mathfrak{m},\mathfrak{m}]$ ones.
For the definitions of these commutators see Eqs.(8)-(9) of Ref.\cite{Antonyan}. The Lie algebraic structure obtained in this way is the one of the largest exceptional group $\mathfrak{e}_8$, and gives rise to the Cartan decomposition.
\beq
\mathfrak{e}_8=\mathfrak{so}(16)\oplus\mathfrak{m}.
\label{Cartendecomp}
\eeq
\noindent
In this picture it is the adjoint representation of $E_8$ which induces on $\mathfrak{m}$ (by restriction) our (\ref{genslocc})
$Spin(16,\mathbb{C})$ action.

Now
using the commutators of the  $[\mathfrak{m},\mathfrak{m}]$ type\cite{Antonyan,Cheng} one can show that the basis vectors of $\overline{\overline{V}}_{S_3}$ are defining a \emph{Cartan subspace} $\mathfrak{c}$ of (i.e. a maximal commutative subspace) of  $\mathfrak{m}$
i.e.
\beq
\overline{\overline{V}}_{S_3}=\mathfrak{c}\subset\mathfrak{m}=\mathcal{F}_8^{+}.
\label{cartan}
\eeq
\noindent
This formula identifies the code subspace as a Cartan subspace of $\mathfrak{e}_8$.
Let moreover $G_0={\rm Spin}(16,\mathbb{C})$. Then it is also known that $G_0\mathfrak{c}$ is a dense subset of 
$\mathcal{F}_8^{+}$.
Hence the generalized SLOCC orbit of our fermionic code is dense 
in the positive chirality subspace of the fermionic Fock space\footnote{Notice also that the subspace of the $[4,2,2]$ four-qubit code, we have started our considerations with, is also dense within the four-qubit space.}.

Let us now consider the state $\vert\Psi\rangle\in \overline{\overline{V}}_{S_3}$ of Eq.(\ref{psi3}). From its amplitudes one can form the $240$ element set of elementary
polynomials $e_s(\Psi_0,\Psi_1,\dots ,\Psi_7), s=1,2,\dots 240$,  of the form
 \beq \pm \Psi_p\pm \Psi_q,\qquad  \frac{1}{2}(\Psi_0\pm \Psi_1\pm \Psi_2\pm\Psi_3\pm \Psi_4\pm \Psi_5\pm \Psi_6\pm \Psi_7),\qquad p\neq q
 \label{e8polinomok}\eeq \noindent where 
 in the second set only an {\it even} number of minus signs are allowed.
This $240=112+128$ split of polynomials corresponds to the root system of the group $E_8$.
 Let us now define the polynomials
 \beq \Pi_{2p}(\Psi) =\sum_s^{240}[e_s(\Psi)]^{2p},\qquad
 2p=2,8,12,14,18,20,24,30\label{Pipol}. \eeq \noindent 
Then it can be shown\cite{Antonyan,LH} that this set is an algebraically independent one of invariant polynomials with respect to the Weyl group
$W\equiv W(\mathfrak{c},\mathfrak{e}_8)$, with $\mathfrak{e}_8$ regarded as a graded algebra. 
It is known\cite{Vinberg} that the restriction of polynomial functions $\mathbb{C}[\mathfrak{m}]\to \mathbb{C}[\mathfrak{c}]$ induces an isomorphism
$\mathbb{C}[\mathfrak{m}]^{G_0}\to\mathbb{C}[\mathfrak{c}]^W$ of the polynomial algebras of $G_0$ and $W$ invariant polynomials.
Since $G_0\mathfrak{c}$ is dense in $\mathcal{F}_8^+$ it follows that any $G_0$ invariant polynomial on $\mathcal{F}_8^+$ is determined by its restriction 
to our fermionic code subspace $\mathfrak{c}$.
In this sense our fermionic code can also be regarded as a seed for generating the algebraically independent set of generalized SLOCC invariant polynomials on $\mathcal{F}_8^+$.

\section{Conclusions}

In this paper we have shown that the $[16,3,4]$ code of Hastings can be obtained  via combining \emph{two} (qubit stabilizer) codes in the  following manner.
As a first step one embeds two copies of the $[4,2,2]$ four qubit code into two different subspaces of the Fock space corresponding to the single and double occupancy representation of four qubits. 
As a result of this one obtains two copies of a $[16,2,4]$ fermionic code.
Then via the use of an intertwiner one glues the four dimensional (two logical qubits) subspaces of these fermionic codes together arriving at the eight dimensional (three logical qubits) one of the $[16,3,4]$ code.
As a byproduct this construction connects our eight dimensional code subspace to a Cartan subspace of $\mathfrak{m}$ answering the Cartan decomposition $\mathfrak{e}_8=\mathfrak{so}(16)\oplus\mathfrak{m}$ of the Lie algebra of $E_8$. 

In arriving at these results we worked out a formalism of embedded qubit systems eligible for fermionic code constructions.
Let us now note in this respect that apart from the single and double occupancy representations of embedded qubits there are mixed occupancy representations as well\cite{LH}.
They are defined as follows.
For the single occupancy subspace let us introduce the notation
\beq
\mathcal{K}_s\equiv \mathcal{K}^{(00\dots 0)}.
\eeq
\noindent
Define a set of $2^n$ dimensional subspaces of $\mathcal{F}_{2n}$ as
\beq
\mathcal{K}^{(\alpha_1\alpha_2\dots\alpha_n)}=c_1^{\alpha_1}c_5^{\alpha_2}\cdots c_{4n-3}^{\alpha_n}\mathcal{K}^{(00\dots 0)},
\qquad (\alpha_1,\alpha_2,\dots\alpha_n)\in\mathbb{Z}_2^n.
\eeq
\noindent
Then one can show that\cite{LH}
\beq
\mathcal{F}_{2n}=\bigoplus_{(\alpha_1\alpha_2\dots\alpha_n)\in\mathbb{Z}_2^n}\mathcal{K}^{(\alpha_1\alpha_2\dots\alpha_n)}.
\eeq
\noindent
Clearly in this notation the double occupancy subspace is
\beq
\mathcal{K}_d\equiv \mathcal{K}^{(11\dots 1)}.
\eeq
\noindent
Now apart from $\mathcal{K}_s$ and $\mathcal{K}_d$ we are having new subspaces. These give rise to mixed occupancy representation of $n$-qubits.
Clearly these considerations define $2^n$ intertwiners similar to the one $\Omega$ showing up in Eq.(\ref{intertwine}).
These operators are the ones responsible for maps in between the subspaces representing embedded $n$-qubit systems.

If in accord to Ref.\cite{Hastings} one wishes to construct fermionic codes for which the chirality operator $\Gamma$ of Eq.(\ref{chirop}) is in the stabilizer:
then one should merely consider subspaces from $\mathcal{F}_{2n}^+$. (Notice that for $n$ odd the single occupancy subspace is not eligible for building such codes. However, one can always use the double occupancy subspace in this respect.)
The next step would be to consider fermionic codes having their origin as qubit stabilizer codes in the spirit of \cite{Terhal}, and examine under what conditions certain number of copies of such codes can be glued together via the use of intertwiners to form new fermionic ones.
Such ideas we are intending to follow in a subsequent publication.

\section{Acknowledgement}
This work was supported by
the Franche-Comt\'e Regional Research Council, Project "Mobilit\'e internationale des chercheurs".
It was also supported by the French "Investissements d'Avenir" program, project ISITE-BFC (contract ANR-15-IDEX-03) and the National Research Development and Innovation Office of Hungary within the Quantum Technology National Excellence Program  (Project No. 2017-1.2.1-NKP-2017-00001)".

\section{Appendix}

Let us elaborate on the structure of the vector space $({\mathbb Z}_2^{2N},\langle\cdot,\cdot\rangle)$ where $\langle\cdot,\cdot\rangle$
is the (\ref{szimpi}) symplectic form. 
Explicitely for the canonical basis $\{e_{\mu}\}$ of ${\mathbb Z}_2^{2N}$ we have
\beq
\mathcal{J}_{\mu\nu}\equiv \langle e_{\mu},e_{\nu}\rangle ={1}_{\mu\nu}-\delta_{\mu\nu},\qquad \mu,\nu=1,2,\dots 2N
\label{szimpimatrix}
\eeq
\noindent 
where ${1}_{\mu\nu}$ is the $2N\times 2N$ matrix having only $1$s as entries.
To the canonical basis vectors we associate the Majorana operators
\beq
e_{\mu}\mapsto c_{\mu},\qquad \mu=1,2,\dots ,2N
\label{basecorr}
\eeq
\noindent
Then
to an arbitrary vector $v=v^{\mu}e_{\mu}\in{\mathbb Z}_2^{2N}$ this mapping assigns a Majorana fermion operator
\beq
v\mapsto c_v\equiv c_1^{v^1}c_2^{v^2}\cdots c_{2N}^{v^{2N}}.
\label{hozzárendel}
\eeq
\noindent
Under this map addition of vectors corresponds to multiplication  of operators modulo elements of
$\omega$ (see Eq.(\ref{omega}).
In this way up to such factors to the $2^{2N}$ vectors of ${\mathbb Z}_2^{2N}$ one can associate $2^{2N}$ Majorana operators $c_{\mathcal{A}}$ where ${\mathcal{A}}$ is one from the $2^{2N}$ possible subsets of $\{1,2,\dots ,2N\}$.
Note, that for a pair of vectors $v=v^{\mu}e_{\mu}$ and $u=u^{\nu}e_{\nu}$ answering the subsets ${\mathcal A}$ and 
${\mathcal B}$ we have 
\beq
\langle v,u\rangle=\mathcal{J}_{\mu\nu}v^{\mu}u^{\nu}
\label{explicitform}
\eeq
\noindent
in accordance with Eq.(\ref{szimpi}).

In (${\mathbb Z}_2^{2N},\langle\cdot,\cdot\rangle)$ one can choose other basis sets dictated by convenience. Consider for instance $\{f_{\mu}\}$ defined by the following set of linear combinations
\beq
f_{2I-1}=e_{2I-1}+e_{2I},\qquad  f_{2I}= \sum_{K=1}^{2I-1}e_{K},\qquad   I=1,2\dots N.
\label{Paulibase}
\eeq
\noindent 
In the Jordan-Wigner representation of our Majorana operators, by virtue of Eqs.(\ref{JW}) and (\ref{basecorr}) 
it is easy to show that our new basis vectors correspond to the ones
\beq
f_{2I-1}\mapsto \sigma_z^{(I)}
\qquad
f_{2J}\mapsto \sigma_x^{(J)},\qquad I=1,2,\dots N.
\label{kanbase}
\eeq
\noindent
In this basis it is easy to check that
\beq
\hat{\mathcal{J}}_{\mu\nu}\equiv \langle f_{\mu},f_{\nu}\rangle,\qquad
\hat{\mathcal{J}}_{2I-1,2I}=\hat{\mathcal{J}}_{2I,2I-1}=1,\qquad \hat{\mathcal{J}}_{\mu\nu}=0,\quad {\rm otherwise}.
\label{convsympl}
\eeq
\noindent
Clearly this property corresponds to the fact that the operators of (\ref{kanbase}) for $I\neq J$ are commuting and for $I=J$ are anticommuting. This basis is usually called the \emph{symplectic basis}.

Writing
\beq
f_{\mu}={e_{\nu}\mathsf{B}^{\nu}}_{\mu},\qquad e_{\nu}=f_{\mu}{\mathsf{A}^{\mu}}_{\nu},\qquad  \mathsf{A}\mathsf{B}=I_{2N}
\eeq
\noindent
with
\beq
\mathsf{B} =
\begin{pmatrix}
1 & 1 & 0 & 1 & \ldots & 1 \\
1 & 0 & 0 & 1 & \ldots & 1 \\
0 & 0 & 1 & 1 & \ldots & 1 \\
0 & 0 & 1 & 0 & \ldots & 1 \\
\vdots & \vdots & \vdots & \vdots & \ddots \\
0 & 0 & 0 & 0 & \ldots & 0
\end{pmatrix},\qquad
\mathsf{A} =
\begin{pmatrix}
0 & 1 & 1 & 1 & \ldots & 1 \\
1 & 1 & 0 & 0 & \ldots & 0 \\
0 & 0 & 0 & 1 & \ldots & 1 \\
0 & 0 & 1 & 1 & \ldots & 0 \\
\vdots & \vdots & \vdots & \vdots & \ddots \\
0 & 0 & 0 & 0 & \ldots & 1
\end{pmatrix}
\end{equation}
by virtue of
$v=v^{\mu}e_{\mu}=\hat{v}^{\nu}f_{\nu}$
we have 
\beq
\hat{v}^{\nu}={\mathsf{A}^{\nu}}_{\mu}v^{\mu}, \qquad {v}^{\nu}={\mathsf{B}^{\nu}}_{\mu}\hat{v}^{\mu}
\label{formula}
\eeq
\noindent
and
\beq
\hat{\mathcal{J}}=\mathsf{B}^T\mathcal{J}\mathsf{B}.
\eeq
\noindent
Noticing that 
\beq
(00)\leftrightarrow I,\quad (01)\leftrightarrow \sigma_x,\quad (11)\leftrightarrow \sigma_y,\quad (10)\leftrightarrow \sigma_z
\eeq
\noindent
one can see that in accordance with (\ref{JW}) the $\mu$ th \emph{column} of $\mathsf{A}$ encodes $c_{\mu}$ expressed in terms of Pauli matrices.
Using Eq.(\ref{formula}) up to a sign any string of Majorana operators can immediately be converted to a one of $N$-fold tensor products of Pauli operators and vice versa.
For example for $N=8$ the generators of the four-qubit code take the following form 
\beq g_j=-\sigma_z^{(2j-1)}\sigma_z^{(2j)},\qquad g_5= \prod_{I=1}^8\sigma_y^{(I)},\qquad g_6=\prod_{j=1}^4\sigma_z^{(2j)}.
\eeq
\noindent
where $j=1,2,3,4$. The chirality operator is $\Gamma=\prod_{I=1}^8\sigma_z^{(I)}$.

\end{document}